\documentclass[a4paper,11pt]{article}
\pdfoutput=1
\usepackage[utf8x]{inputenc}
\usepackage{jheppub}
\usepackage{epsf}
\usepackage{color}
\usepackage{subfigure}
\usepackage{ulem}
\usepackage{slashed}
\newcommand{\lag}{\mathcal{L}}

\newcommand{\eea}{\end{eqnarray}}
\newcommand{\eq}[1]{Eq.~(\ref{#1})}

\newcommand{\nn}{\nonumber}

\def\stw{s_{\theta_W}}
\def\ctw{c_{\theta_W}}
\def\ttw{t_{\theta_W}}
\def\bea{\begin{eqnarray}}
\def\eea{\end{eqnarray}}
\def\lra#1{\overset{\text{\scriptsize$\leftrightarrow$}}{#1}}
\usepackage{array}
\newcolumntype{C}[1]{>{\centering\let\newline\\\arraybackslash\hspace{0pt}}m{#1}}

\makeatletter
\def\section{\@startsection {section}{1}{\z@}{-3.5ex plus -1ex minus
 -.2ex}{2.3ex plus .2ex}{\large\bf}}
\def\subsection{\@startsection{subsection}{2}{\z@}{-3.25ex plus -1ex
minus -.2ex}{1.5ex plus .2ex}{\normalsize\bf}}
\makeatother
\makeatletter

\@addtoreset{equation}{section}

\makeatother

\begin{document}

\title{Running away from  the $T$-parameter  solution to the $W$-mass anomaly}
\author{Rick S. Gupta}
\affiliation{Department of Theoretical Physics,
Tata Institute of Fundamental Research, Homi Bhabha Rd, Mumbai 400005, India}

\date{\today}

\abstract{We show that it is essential to include renormalisation group (RG) effects  for  determining the SMEFT parameter space consistent with the CDF $W$-mass anomaly at the matching scale. This is because operators that are only weakly constrained--for instance those probed by diboson and Higgs data--can have a contribution to the $W$-boson mass via one-loop RG effects, that is comparable to the tree-level contribution of the much more strongly constrained operators related to electroweak precision observables. We RG evolve the low energy SMEFT parameter space consistent with the anomaly to the matching scale and find a much larger allowed region. In particular we find that it is possible to have a vanishing or even negative $T$-parameter at the matching scale. This will hopefully  lead to a larger set of UV completions that can explain the anomaly. For the one-loop contributions to be important a relatively low new physics scale around 800 GeV is required. This enhances the possibility of probing this new physics in direct and indirect searches in the recent future.}

\maketitle

\section{Introduction}
\label{intro}

The CDF collaboration has announced the most precise measurement ever  of the $W$-boson mass~\cite{CDF},
\bea
m^{\rm CDF}_W =(80.433\pm 0.0064_{\rm stat}\pm 0.0069_{\rm syst}) {\rm~GeV}, 
\eea
This measurement disagrees with the Standard Model (SM) prediction, 
\bea
m_W = (80.357±0.004_{\rm inputs}\pm 0.004_{\rm theory}) {\rm~GeV}
\eea
at the 7$\sigma$ level and  also with the  global average of previous measurements~\cite{pdg}, 
\bea
m_W =(80.379 ± 0.012) {\rm~GeV}.
\eea

This has led to  multiple  proposals~\cite{Bagnaschi:2022whn, Strumia, Balkin:2022glu, Fan:2022yly, deBlas:2022hdk, Asadi:2022xiy, Lu:2022bgw, Carpenter:2022oyg,Fan:2022dck, Zhu:2022tpr, Zhu:2022scj, Kawamura:2022uft, Nagao:2022oin, Zhang:2022nnh, Liu:2022jdq, Sakurai:2022hwh, Cacci, Song:2022xts, Bahl:2022xzi, Cheng:2022jyi, Babu:2022pdn, Heo:2022dey, Ahn:2022xeq, Zheng:2022irz, Perez:2022uil, Kanemura:2022ahw, Borah:2022obi, Popov:2022ldh, Arcadi:2022dmt, Ghorbani:2022vtv,Han:2022juu, Du:2022pbp, Tang:2022pxh,Yang:2022gvz, Athron:2022isz,Ghoshal:2022vzo, Athron:2022qpo, Blennow:2022yfm,Heckman:2022the, Lee:2022nqz, DiLuzio:2022xns,Paul:2022dds,Biekotter:2022abc,Cheung:2022zsb,Du:2022brr, Endo:2022kiw, Crivellin:2022fdf, Mondal:2022xdy, Chowdhury:2022moc, Du:2022fqv, Bhaskar:2022vgk, Yuan:2022cpw, Arias-Aragon:2022ats}  exploring a BSM solution to this discrepancy.   Many of the proposed solutions require heavy new physics making effective field theories (EFT) an ideal approach to tackle this issue.  A two-step methodology  has been frequently utilised   in these works:  (1) new  Standard Model Effective Theory (SMEFT) operators are introduced to resolve the anomaly~\cite{ Bagnaschi:2022whn, Strumia, Balkin:2022glu, Fan:2022yly, deBlas:2022hdk, Asadi:2022xiy, Lu:2022bgw, Carpenter:2022oyg}, followed by    (2) an exploration of UV complete models that can  generate the observed low energy pattern of operators. In particular the introduction of operators that induce the Peskin-Takeuchi $(S, T)$-parameters--or indeed even just a positive value for the $T$-parameter--can provide  a very good  fit to the data. In fact, irrespective of the value of the other Wilson coefficients,  it is difficult  to obtain a very good fit to the data  without some positive $T$ contribution (see for instance Ref.~\cite{Bagnaschi:2022whn, Strumia, Balkin:2022glu}). This has naturally led to an effort to identify models that generate such a  positive $T$ contribution  and resolve the anomaly.

In this work we show that a crucial intermediate step has been missed in the above analysis: that of  evolving the constraints on the Wilson coefficients from the $m_W$-scale to the matching scale using the renormalisation group equations (RGE) for the Wilson coefficients.  We will show that, at the matching scale, this significantly enhances the SMEFT  parameter space   capable of resolving the anomaly.  As we will show the SMEFT parameter space  explored so far as a solution to this anomaly is in fact just a  subset of a much larger region.  In particular, we will find  that  it is possible to obtain a very good fit to the data even if, ${\cal O}_T=\frac{1}{2}\left (H^\dagger {\lra{D}_\mu} H\right)^2$, the operator that generates the $T$-parameter, vanishes at the matching scale

It might seem  surprising how one-loop suppressed RG effects can have  a major impact especially given the short span of energy scales over which the Wilson coefficients  run in this case.  As it was pointed out in Ref.~\cite{GuptaU}, however, RG effects can  become  important because all the operators in the electroweak sector have not been constrained at the same level of precision. Thus a poorly constrained operator with a relatively large Wilson coefficient can have a significant contribution to a very strongly bounded operator despite  the loop suppression. In particular we will show that the RG contributions of operators--that are only weakly constrained by diboson and Higgs data--to the strongly constrained operators generating  $S$ and $T$,  can be of the same order as the tree level contribution of the latter operators.\footnote{ In the context of  the electroweak chiral lagrangian,  it was already observed in Ref.~\cite{Dutta}  that the running of operators contributing to diboson production  can have a significant impact on the allowed parameter space for the $S$-$T$ parameters at  a high scale.} As these RG effects  are one loop suppressed, a resolution of the anomaly requires a relatively low scale ($\Lambda\sim800$ GeV) for the contributing operators which in turn implies sizeable anomalous Higgs and  triple gauge couplings  (TGC). This gives hope that if  new physics  is responsible for this anomaly, it  can be probed both in direct and indirect searches in the coming years.

\section{RG induced contributions to  the $W$-mass}

In this section we show the importance of RG effects in deriving the SMEFT parameter space consistent with the CDF $W$-mass anomaly. We will show this in detail for the case of universal new physics and then argue that our main observations will hold also for the general case.  We will first discuss in Sec.~\ref{bounds} the relative level of constraints on the different universal operators from electroweak precision observables, TGC bounds and Higgs physics. This will then let us identify the weakly constrained operators that can have a large impact on the RG flow  of Wilson coefficients that contribute to the $W$-mass in Sec.~\ref{RGsec}. Finally in Sec.~\ref{nuniv} we will discuss how this analysis can be extended beyond the universal case.

\subsection{Operators  for universal new physics and current constraints} 
\label{bounds}

\begin{table}
\small
\centering
\begin{tabular}{ccc}
\begin{tabular}{|c|}\hline
${\cal O}_H=\frac{1}{2}(\partial^\mu |H|^2)^2$\\
${\cal O}_T=\frac{1}{2}\left (H^\dagger {\lra{D}_\mu} H\right)^2$\\
${\cal O}_r=|H|^2 |D_\mu H]^2$\\
${\cal O}_{K4}=|D^2 H|^2$\\
${\cal O}_6=\lambda |H|^6$ \\
${\cal O}_{HW}=ig  ({D^\mu}  H)^\dagger  \sigma^a {D^\nu} H D^\nu  W_{\mu \nu}^a$\\
${\cal O}_{HB}=ig' ({D^\mu}  H)^\dagger  {D^\nu} \partial^\nu  B_{\mu \nu}$\\
\hline 
\end{tabular}
&\qquad \qquad&
\begin{tabular}{|c|}\hline
${\cal O}_{2W}=-\frac{1}{2}  ( D^\mu  W_{\mu \nu}^a)^2$\\

${\cal O}_{2B}=-\frac{1}{2}( \partial^\mu  B_{\mu \nu})^2$\\
${\cal O}_{2G}=-\frac{1}{2}  ( D^\mu  G_{\mu \nu}^A)^2$ \\ 
${\cal O}_{BB}=g^{\prime 2} |H|^2 B_{\mu\nu}B^{\mu\nu}$\\
${\cal O}_{WB}= g{g}^{\prime} H^\dagger \sigma^a H W^a_{\mu\nu}B^{\mu\nu}$\\
${\cal O}_{WW}=g^2 |H|^2 W^a_{\mu\nu}W^{a \mu\nu}$\\
${\cal O}_{GG}=g_s^2 |H|^2 G_{\mu\nu}^A G^{A\mu\nu}$\\
${\cal O}_{3W}= \frac{1}{3!} g\epsilon_{abc}W^{a\, \nu}_{\mu}W^{b}_{\nu\rho}W^{c\, \rho\mu}$\\
${\cal O}_{3G}= \frac{1}{3!} g_s f_{ABC}G^{A\, \nu}_{\mu}G^{B}_{\nu\rho}G^{C\, \rho\mu}$
 \\\hline
 \end{tabular}
\end{tabular}
\caption{The 16 CP-even operators that can give a general parametrisation of universal new physics effects. Here,  $D_{\rho} W^a_{\mu \nu} = \partial_\rho W^a_{\mu \nu} + g \epsilon^{abc} W^b_\rho W^c_{\mu\nu}$, $H^\dagger {\lra { D_\mu}} H\equiv H^\dagger D_\mu H - (D_\mu H)^\dagger H $, with $D_\mu H = \partial_\mu H -i g\tau^a W^a_\mu H - i g' Y_H B_\mu H$ and $\sigma^a$ are the Pauli matrices.
 \label{table1}}
\end{table}

In `universal' theories, as defined in Ref.~\cite{wells}, the 16 bosonic operators in Table~\ref{table1} can be used for a general parametrisation of the low energy theory.  Not all of  these operators would enter of our analysis. In order to be   relevant for our analysis, an operator should either, (1) contribute to $m_W$ at tree level or via a one loop RG contribution, or, (2) it should be important for extracting constraints on the operators of the first category. This gives us the following set of operators,
\bea
\{{\cal O}_H, \ {\cal O}_T, \ {\cal O}_r, \ {\cal O}_{K4},\ {\cal O}_{HW}, \ {\cal O}_{HB}, \ {\cal O}_{2W}, \ {\cal O}_{2B}, \ {\cal O}_{GG}, \ {\cal O}_{WW}, \ {\cal O}_{WB}, \ {\cal O}_{BB}, {\cal O}_{3W}\}.
\label{set10}
\nonumber
\eea
Data from electroweak precision observables, Higgs physics and diboson data from  LEP and LHC, can be used to constrain the operators in \eq{set10}.  We have used the so-called Hagiwara-Ishihara-Szalapski-Zeppenfeld (HISZ) basis~\cite{hisz} where  it is especially easy to separate the less constrained operators--which are probed  by Higgs and diboson processes-- from those contributing to the precisely measured electroweak precision observables.

\begin{figure}[t]
        \centering
\includegraphics[scale=0.75]{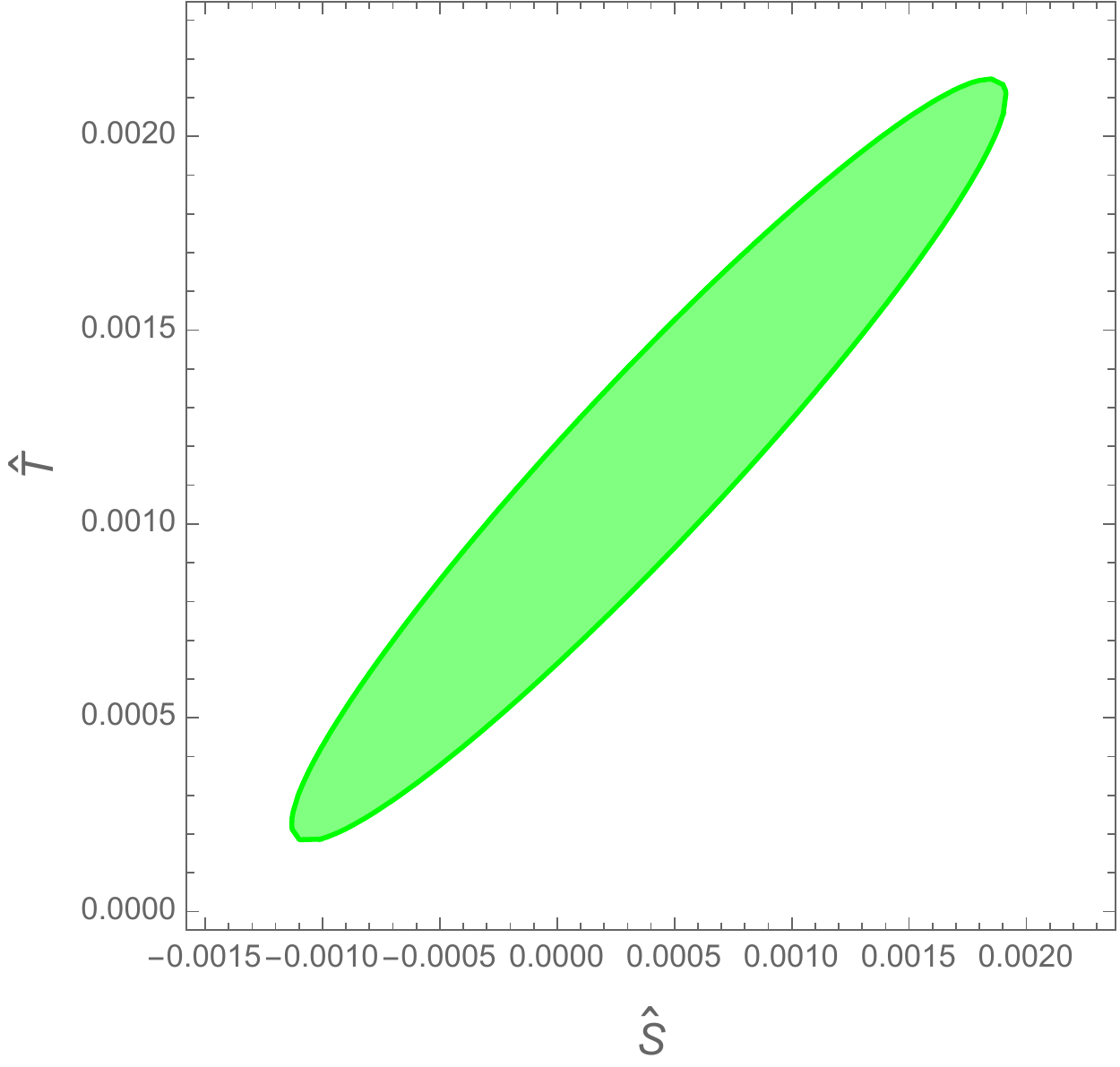}
\caption{ The allowed region in the $\hat{S}$-$\hat{T}$ plane, as derived in Ref.~\cite{Strumia}, after taking into account the new CDF $W$-mass measurement and marginalising over $W$ and $Y$.}
 \label{green}
 \end{figure}

  Let us now discuss the constraints on these operators. We start with the  electroweak precision observables which include the  $Z$-pole measurements at LEP 1,  observables from the  $e^+e^- \to f^+f^-$ process at LEP-2 and $pp \to l \bar{l}$ at LHC, and indeed the   $W$-mass. Universal  corrections to these observables up to the dimension 6 level   can be parametrised by the $\hat S$, $\hat T$, $W$ and $Y$ parameters~\cite{Peskin, Barbieri}. In the $(\alpha_{em},G_F, m_Z)$ input scheme the contributions of the operators in \eq{set10} to these parameters is given by,
 \bea
&& \hat{T}= \frac{v^2}{\Lambda^2} c_T(m_W) \ , \quad 
	\hat{S}=  \frac{g^2 v^2}{4\Lambda^2} 4 c_{WB}(m_W), \nonumber\\
&& Y=\frac{g^2 v^2}{4\Lambda^2} c_{2B}(p) \ , \qquad
	W = \frac{g^2 v^2}{4\Lambda^2} c_{2W}(p) \  . 
 \label{EWPT} 
 \eea
where $\hat{S}={\alpha_{em}} S/4 s^2_{\theta_W}$, $\hat{T}=\alpha_{em} T$, $S$ and $T$ being  the original definitions of Ref.~\cite{Peskin}.   Ref.~\cite{Strumia} presents the updated electroweak fit including these 4 parameters and assuming the CDF measurement of the $W$-mass is correct. Whereas the introduction of only the $W$ and $Y$  parameters is unable  to provide a good fit once the CDF $W$-mass measurement is included, allowing for a non-vanishing  $\hat{S}$ and $\hat{T}$ (or indeed  only $\hat{T}$) can provide a very good fit to the data. Data from LEP-2 and especially  high energy processes at the LHC strongly constrains the parameters $|W|, |Y|\lesssim 10^{-4}$~\cite{Strumia}; the energy scale, $p$, in \eq{EWPT} corresponds to the energy scale of the process used to constrain $W$ and $Y$. We reproduce in Fig.~\ref{green}  the allowed region in the $\hat{S}$-$\hat{T}$ plane  from Ref.~\cite{Strumia}, which assumes that the new CDF $W$-mass measurement is correct and marginalises over $W$ and $Y$.

\begin{figure}
        \centering
 \includegraphics[scale=0.55]{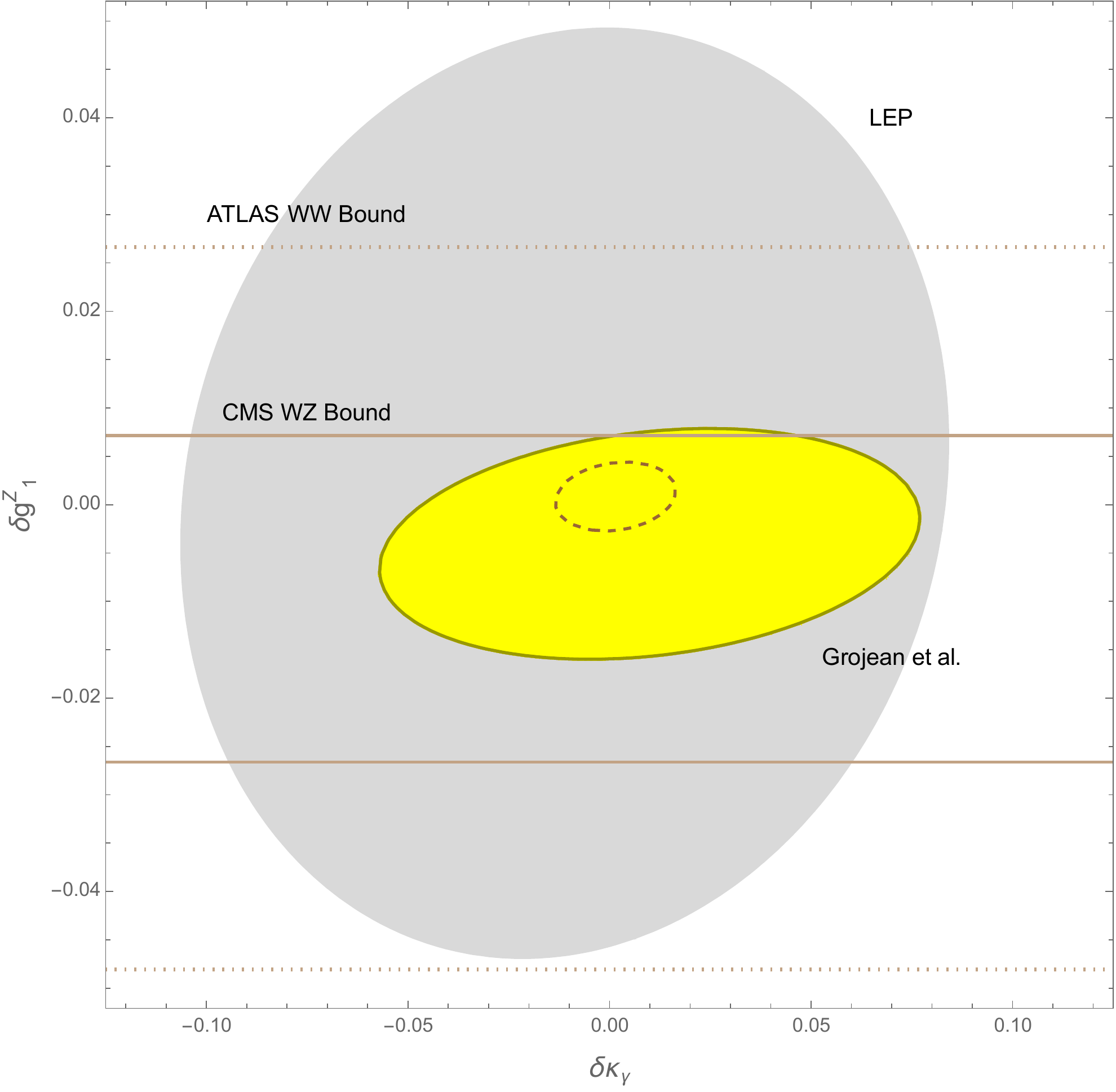}
\caption{  LEP~\cite{lep} and LHC~\cite{GrojeanM}  bounds  in the  $(\delta \kappa_\gamma,  \delta g_1^Z)$ plane after profiling over $\lambda_\gamma$. The dashed curve shows the HL-LHC projection from Ref.~\cite{GrojeanM} . We also show the bounds reported by ATLAS from the $WW$ process~\cite{atlasww} and by CMS from the $WZ$  process~\cite{cmswz}.
}
 \label{yellow}
 \end{figure}

Next we come to LEP-2 and LHC diboson data that  can probe modifications of the SM $WWZ$ and $WW\gamma$ vertices, the so called anomalous  Triple Gauge Couplings (TGC) defined in Ref.~\cite{hagZ}. While the operators generating TGC do not contribute to the $W$-mass at tree level they do contribute to it via one loop RG effects as we will see in the next subsection. In terms of the Wilson coefficients of these operators the TGC are given by,
\bea \begin{split}
	& \delta g_1^Z = - \frac{g^2 v^2}{4\Lambda^2} \frac{1}{c_{\theta_W}^2} c_{HW} (p) \ , \qquad
	\delta \kappa_\gamma = -\frac{g^2 v^2}{4\Lambda^2} (c_{HW}(p)+c_{HB}(p)-4 c_{WB}(p)) \ , \\
	& \lambda_\gamma =  \frac{g^2 v^2}{4\Lambda^2} c_{3W}(p).
\end{split}  
\label{TGC} 
 \eea
 In Fig.~\ref{yellow} we show the LEP~\cite{lep} and much stronger LHC bounds on these TGC in the  $(\delta \kappa_\gamma,  \delta g_1^Z)$ plane after profiling over $\lambda_\gamma$. For the LHC bounds  shown in yellow we have used the analysis of Ref.~\cite{GrojeanM}, which also constrains $|\lambda_\gamma]\lesssim0.01$. The dahed curve shows the HL-LHC projections from Ref.~\cite{GrojeanM}. We also show,  in Fig.~\ref{yellow},  the  bounds reported by ATLAS from the $WW$ process~\cite{atlasww} and by CMS from the $WZ$  process~\cite{cmswz}.   In these LHC studies, the high energy part of  differential distributions   is used to derive strong bounds on the TGC so that the energy scale, $p$, in \eq{TGC}   is not $2 m_W$ but the energy scale of the  events being used to derive the bound.  One can thus directly read off the high scale values of the Wilson coefficients from the bounds in Fig.~\ref{yellow}.

Finally we discuss the bounds on the operators $\{{\cal O}_H, \ {\cal O}_r, \ {\cal O}_{K4},\ {\cal O}_{WW},\ {\cal O}_{BB}\}$ which can be constrained  by data from  Higgs Physics. Again these operators have no direct impact on the electroweak fit that yields Fig.~\ref{green}, but will become important in the RG analysis of Sec.~\ref{RGsec}. First, it will be convenient to rewrite the operators, $\{{\cal O}_H, \ {\cal O}_r \}$, as follows, 
\bea
\frac{c_H}{\Lambda^2}{\cal O}_H+\frac{c_r}{\Lambda^2}{\cal O}_r=\frac{c_{H+}}{\Lambda^2}{\cal O}_{H+}+\frac{c_{H-}}{\Lambda^2}{\cal O}_{H-}
\label{rew}
\eea
where ${\cal O}_{H\pm}=\frac{1}{2}({\cal O}_H \pm {\cal O}_r)$ and $c_{H\pm}=c_H \pm c_r$. The operators ${\cal O}_{H\pm}$ and ${\cal O}_{K4}$ when expanded generate the following anomalous couplings, 
\bea
	\Delta \lag_{H} &=& \delta \kappa_V \frac{g^2 v}{2}h\left(W^{+}_\mu W^{-\mu}+\frac{Z_\mu Z^\mu}{2 c_{\theta_W}}\right)-\delta \kappa_f \sum_f \frac{m_f}{v}h \bar{f} f+\frac{\kappa_{gg} h}{2 v}{G}_{\mu \nu}{G}^{\mu \nu}\nonumber\\&+&\frac{\kappa_{\gamma\gamma} h}{2 v}{A}_{\mu \nu}{A}^{\mu \nu}+ \frac{\kappa_{\gamma Z}~h}{v} {A}_{\mu \nu}{Z}^{\mu \nu}, \nonumber
	\label{higgsL}
\eea
which can be determined in a fit to Higgs data. The five  operators,
\bea
  \{{\cal O}_{H+}, \ {\cal O}_{H-}, \ {\cal O}_{K4},\ {\cal O}_{WW},\ {\cal O}_{BB}\}
  \eea
  contribute to these anomalous couplings as follows,
\bea 
\delta \kappa_V&=&-\frac{c_{H-}(m_h) ~v^2}{2\Lambda^2}\nn\\
\delta \kappa_f&=&\left(-c_{H+}(m_h)- 2 \lambda c_{K4}(m_h)\right) \frac{v^2}{2\Lambda^2}\nn\\
	{\kappa_{\gamma\gamma}} &=& 2 g^2 \stw^2\left(c_{WW}(m_h)+   c_{BB}(m_h)-    c_{WB}(m_h)\right) \frac{v^2}{\Lambda^2} \nn\\
{\kappa_{Z \gamma}} &=&g^2 \ttw \left( 2 \ctw^2 c_{WW}(m_h)-2 \stw^2 c_{BB}(m_h)-  (\ctw^2- \stw^2) c_{WB}(m_h) \right)\frac{v^2}{\Lambda^2}\nn\\
{\kappa_{GG}} &=&2 g_s^2 \frac{ c_{GG}(m_h)~v^2}{\Lambda^2}  
 \eea
We see that our rewriting of operators in \eq{rew} has allowed us to identify the linear combination, ${\cal O}_{H-}$,  of the operators ${\cal O}_H$ and ${\cal O}_r$ that modifies the gauge-Higgs vertex.  It is this same linear combination that gives  RG contributions to $S$ and $T$ as we will see in Sec.~ \ref{RGsec}.    In Ref.~\cite{atlasfit} a combined fit yields the following 95$\%$ CL bounds,  
\bea
-0.03&\leq&\delta  \kappa_V\leq 0.13\nn\\
-0.13&\leq&\delta  \kappa_f\leq 0.23,
\label{kappalimit}
\eea
which implies only weak bounds on $c_{H-}, c_{H+}$ and $c_{K4}$. This leaves the  couplings $\kappa_{\gamma\gamma}, ~\kappa_{Z\gamma}$ and $\kappa_{GG}$ which can all be constrained very strongly to be smaller than ${\cal O}(10^{-3})$. The reason for these strong bounds is that the  $h \to \gamma \gamma, h\to Z \gamma$ and $gg \to h$ processes are loop induced so that even a small  change in these couplings can cause a large fractional change in their rates~\cite{priva}. Given that $c_{WB}$ is very precisely constrained due to its contribution to $\hat{S}$, this implies that all of the three Wilson coefficients, $c_{GG}, c_{BB}$ and $c_{WW}$, can be very strongly  constrained. 

To summarise, the operators, $\{{\cal O}_T,  \ {\cal O}_{WB}, \ {\cal O}_{2W}, \ {\cal O}_{2B}\}$, can be very precisely constrained by the electroweak fit whereas the operators  $\{ {\cal O}_{GG}, \ {\cal O}_{WW}, \ {\cal O}_{BB}\}$ are strongly bounded by the  $h \to \gamma \gamma, h\to Z \gamma$ and $gg \to h$ processes. Next, the set of operators, $\{{\cal O}_{HB},  \ {\cal O}_{HW}, \ {\cal O}_{3W} \}$,  are constrained at a much weaker level by their contributions to the  TGC. Finally the operators,   $\{{\cal O}_{H+}, \ {\cal O}_{H-}, \ {\cal O}_{K4}\}$  are constrained even more weakly from Higgs data.

\subsection{RG evolution of constraints to matching scale} 
 \label{RGsec}

%

\begin{figure}
        \centering
        \begin{tabular}{ccc}
             \includegraphics[scale=0.37]{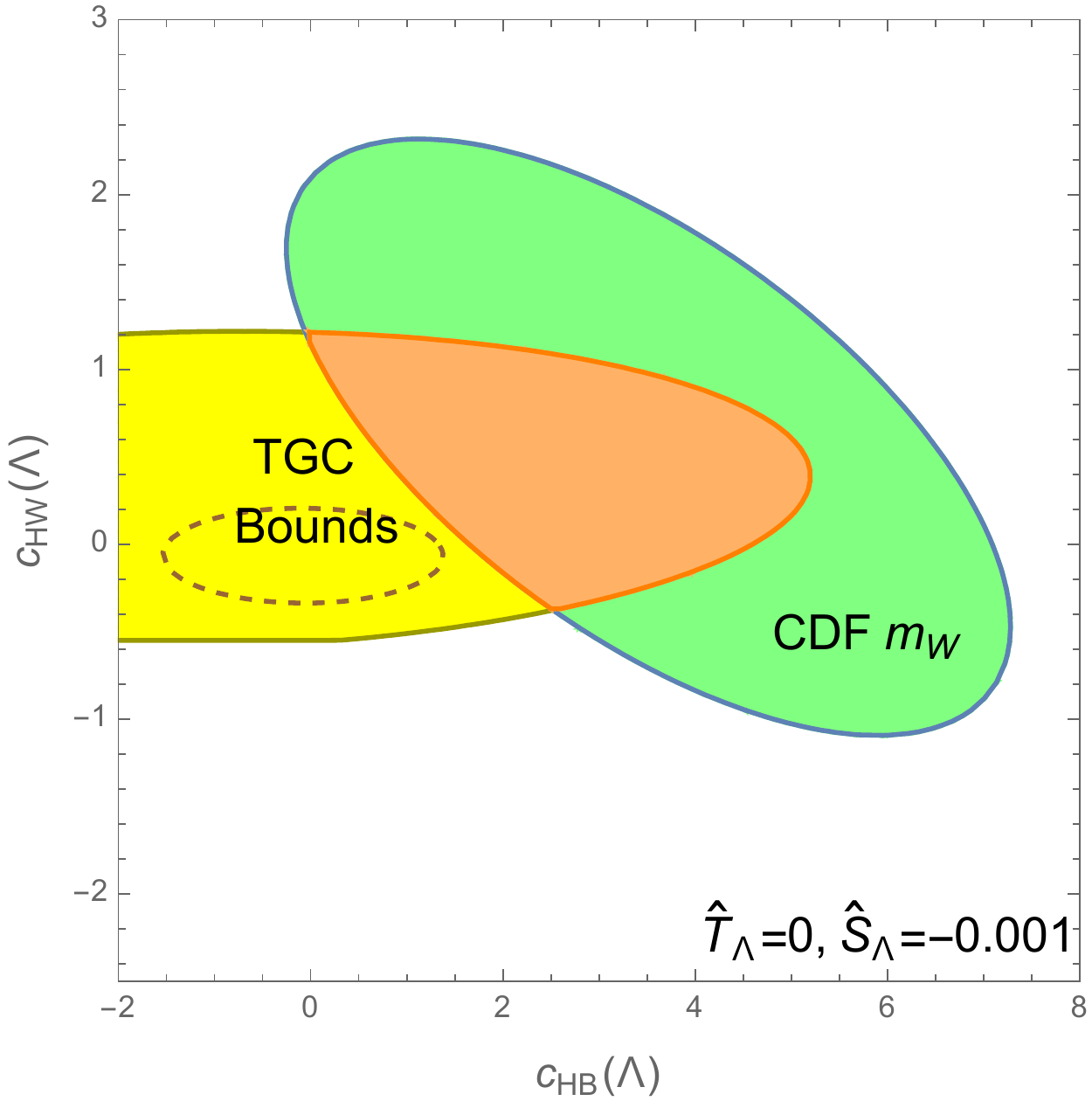}& \includegraphics[scale=0.37]{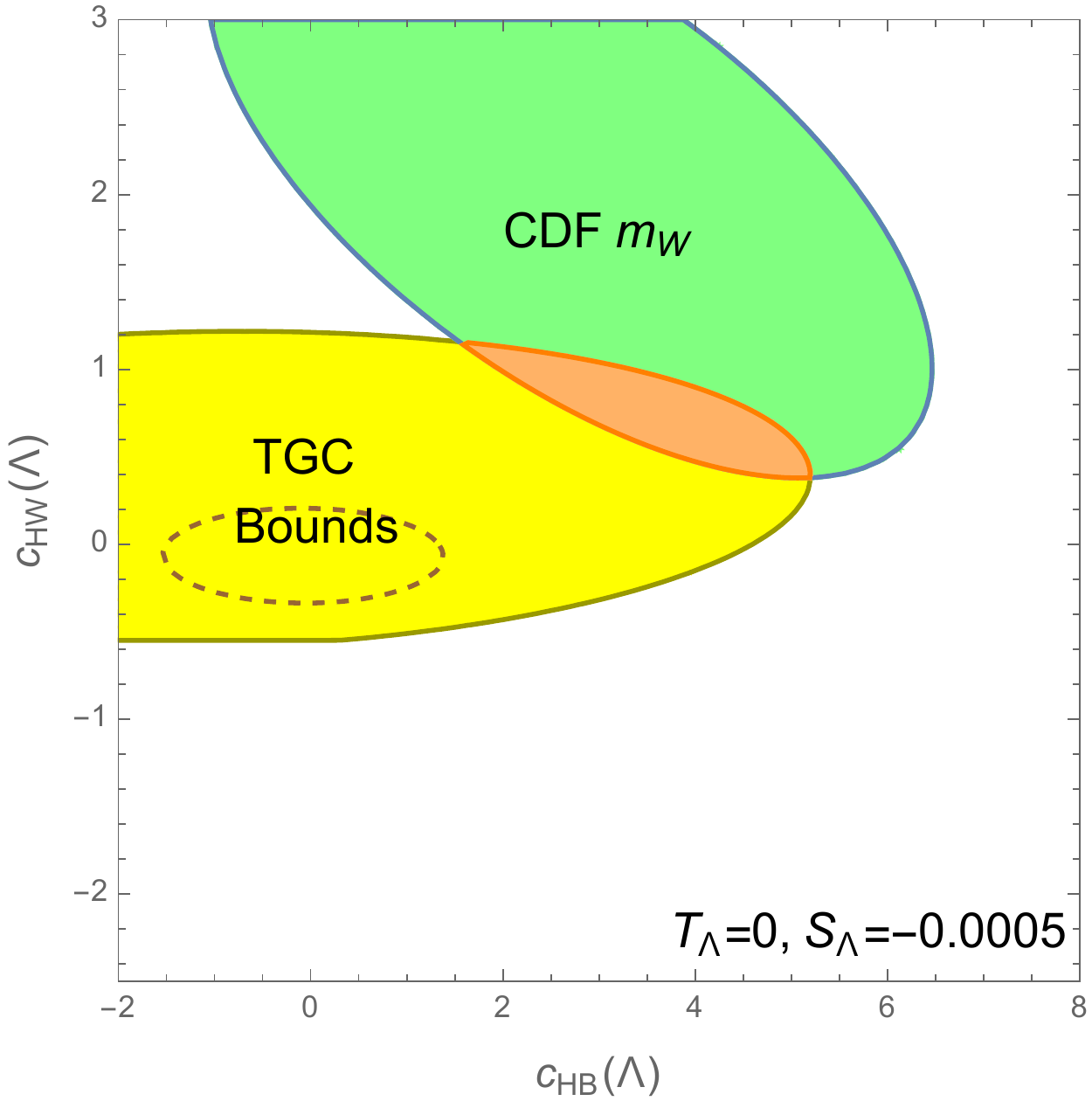} &\includegraphics[scale=0.37]{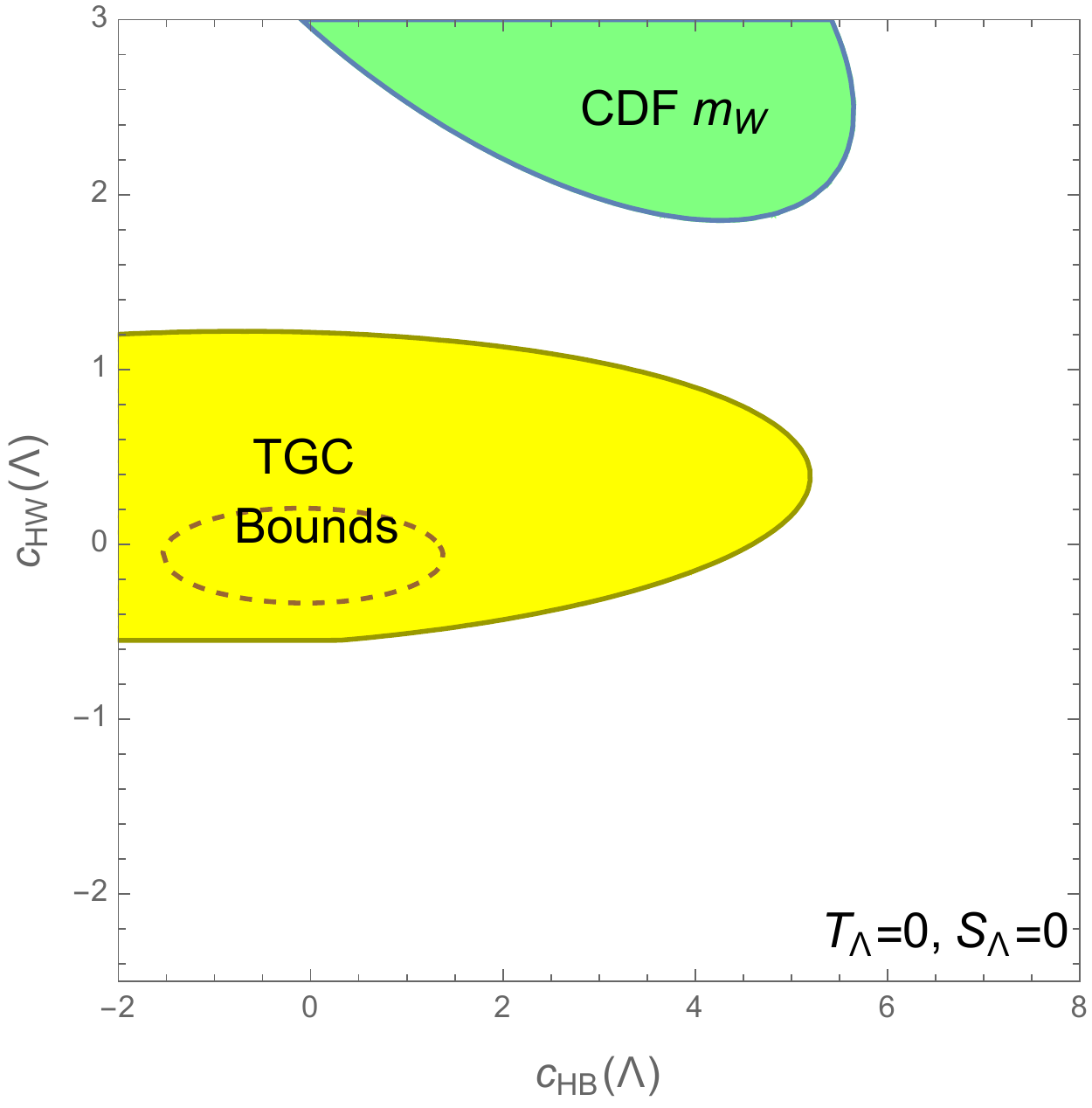} \\
             \includegraphics[scale=0.37]{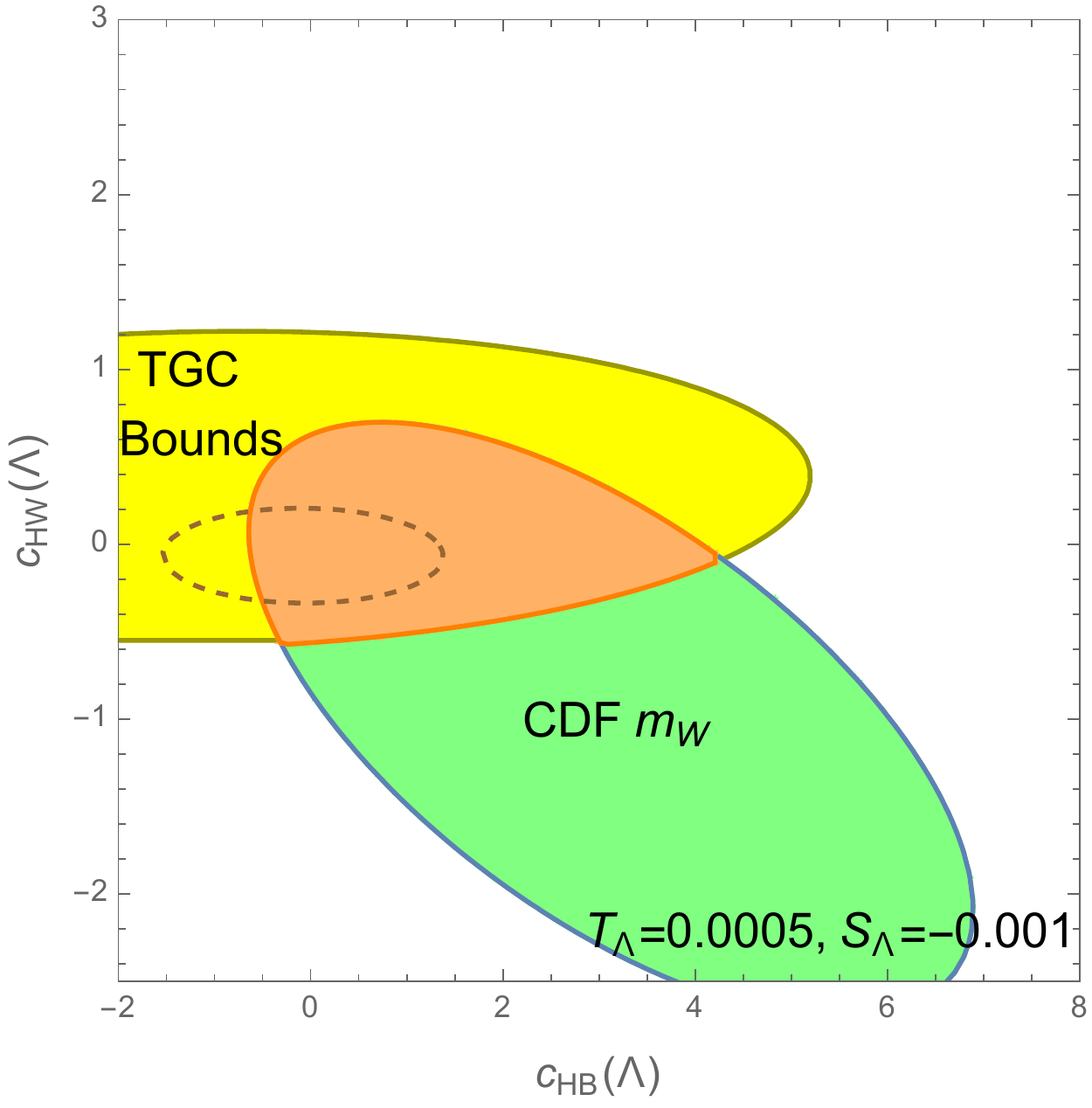} & \includegraphics[scale=0.37]{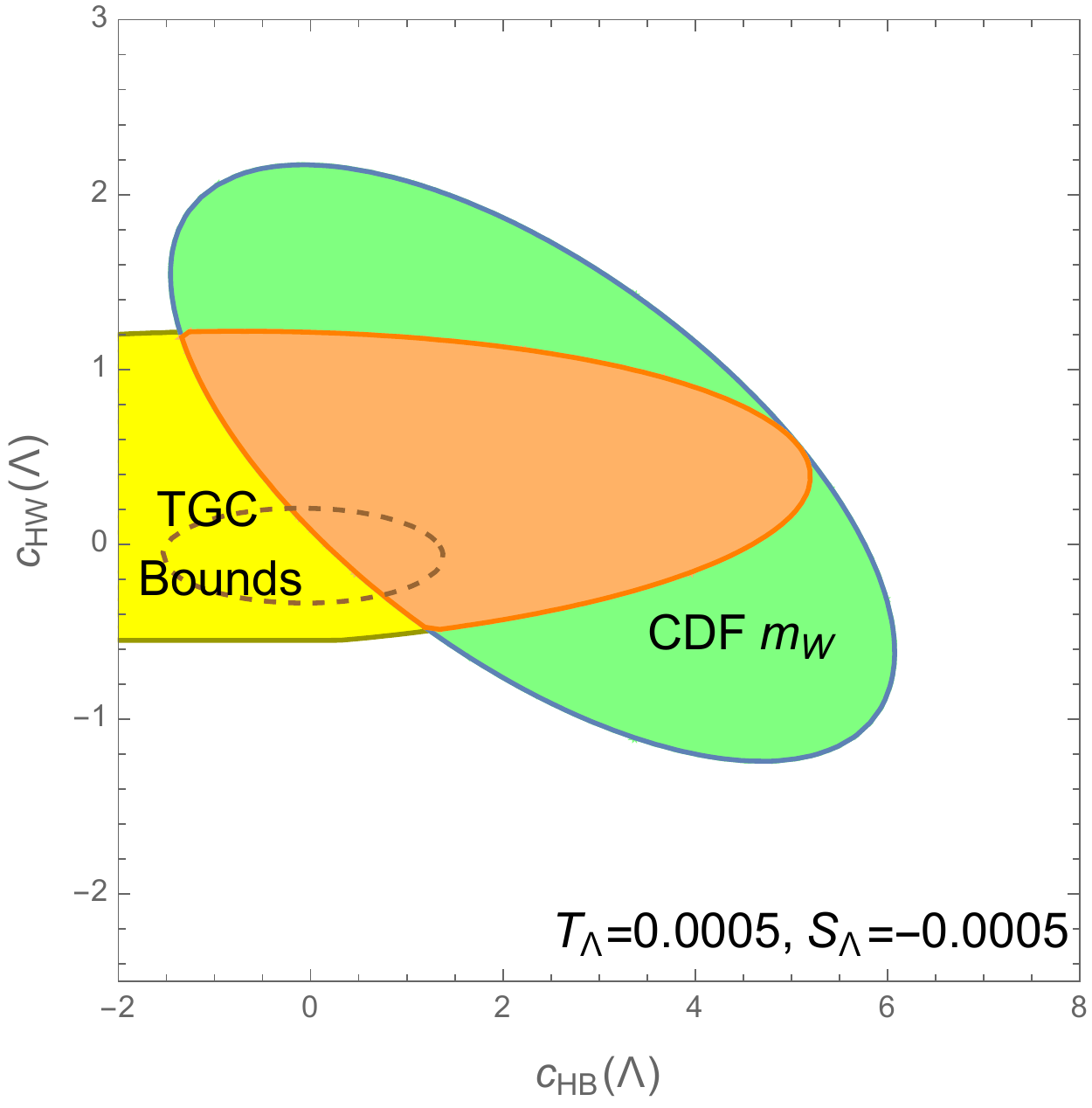}&\includegraphics[scale=0.37]{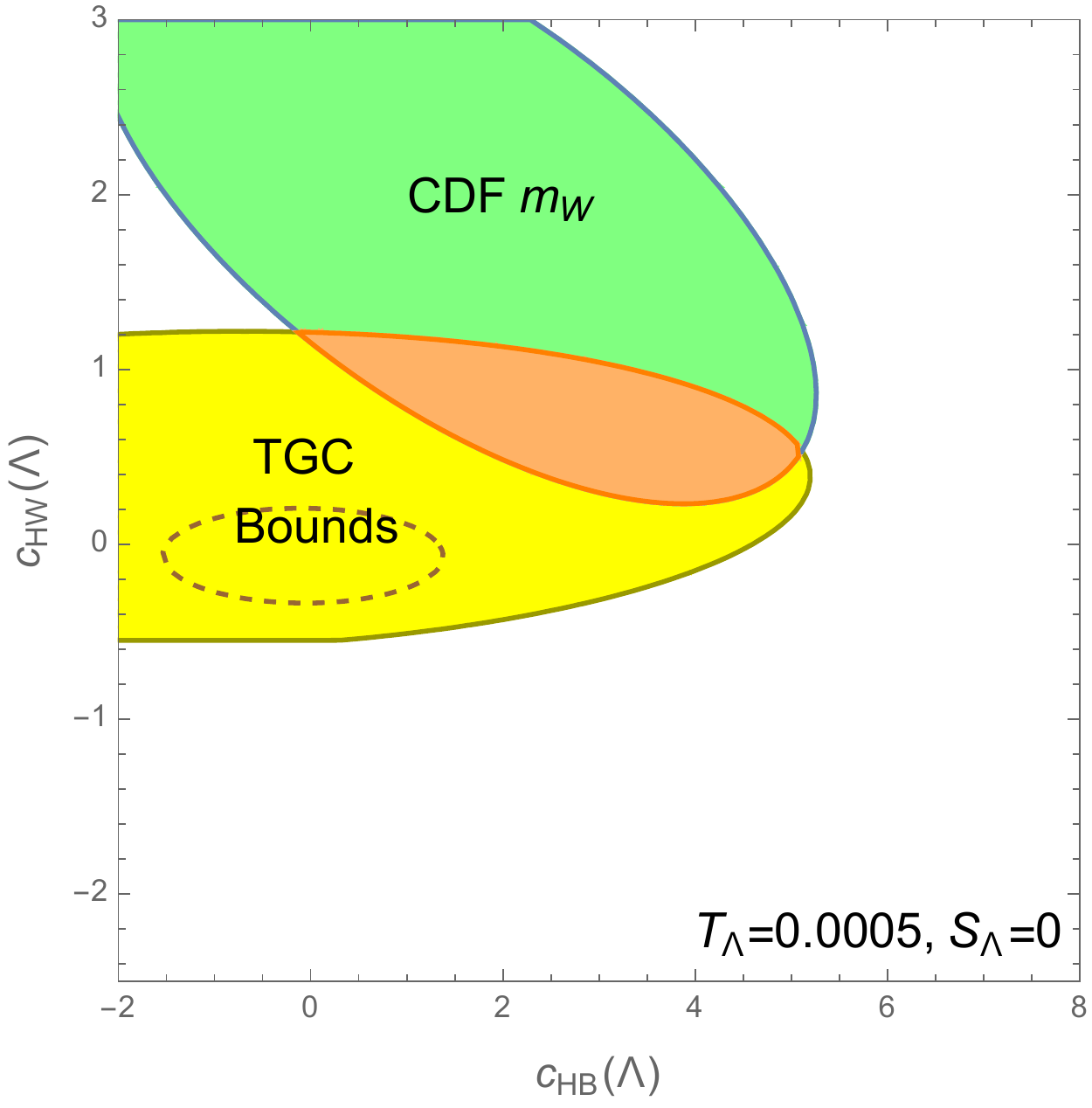}  \\
                  \end{tabular}
        \caption{We show in green slices of the 4 dimensional region consistent with the CDF $m_W$ anomaly in $c_{HW}(\Lambda), c_{WB}(\Lambda)$ planes for fixed values of $ \hat{S}_\Lambda$ and $ \hat{T}_\Lambda$. The TGC bounds from Fig.~\ref{yellow} are shown in yellow. We show the TGC sensitivity projections from Ref.~\cite{GrojeanM} by the dashed curve. A explanation for the CDF anomaly exists only if there is an overlap between the green and yellow regions. We have taken $\Lambda=10~m_W$.}
        \label{slices}
    \end{figure}

 \begin{figure}
 \centering
\includegraphics[scale=0.8]{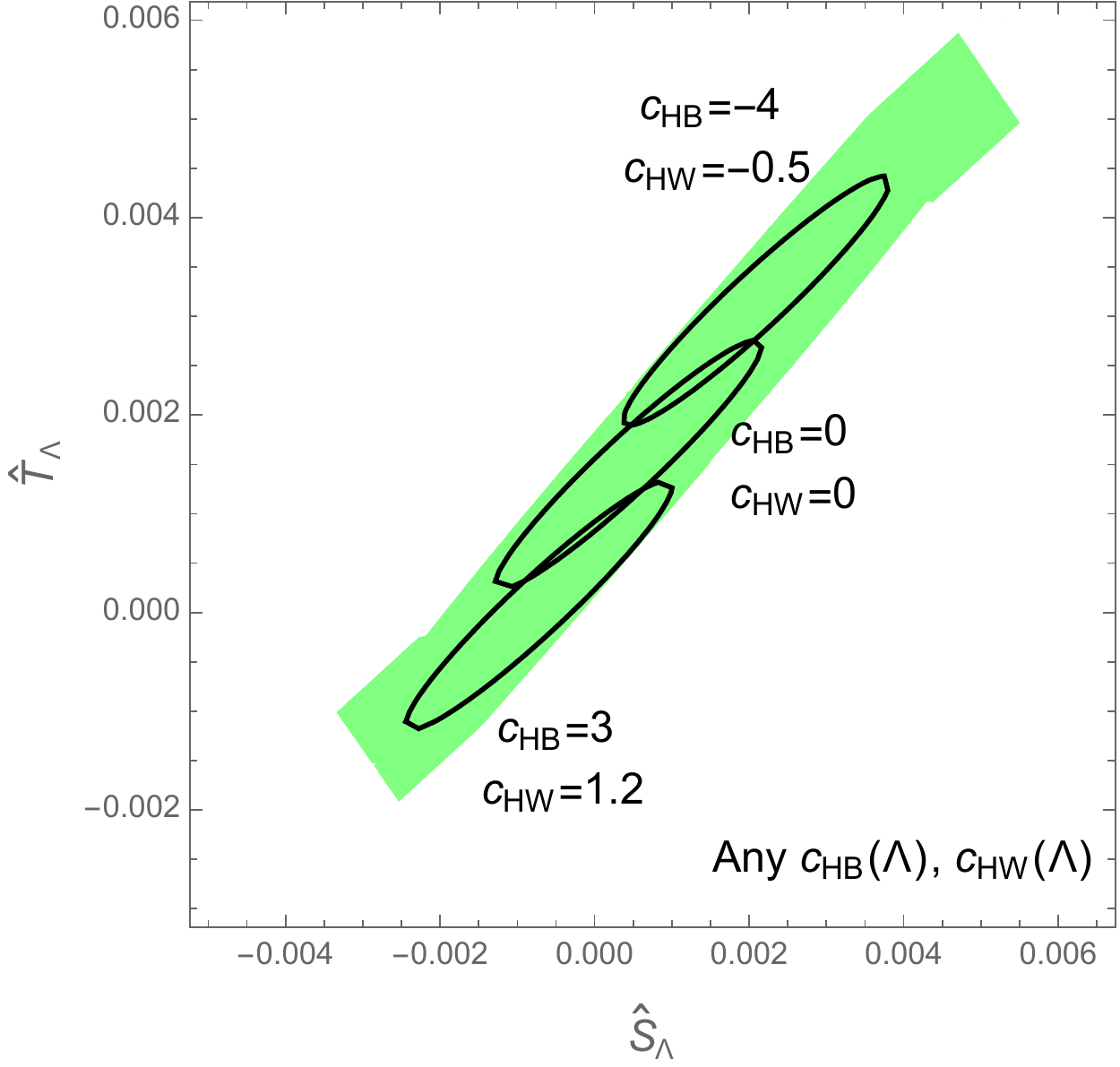}
\caption{ The region is in green consistent with the CDF anomaly after marginalising over $c_{HB}(\Lambda)$ and $c_{HW}(\Lambda)$  while respecting the TGC bounds from Fig.~ \ref{yellow}. The black curves show the bounds for the fixed values  $c_{HB}(\Lambda)$ and $c_{HW}(\Lambda)$ that have been labelled in the plot. We have taken $\Lambda=10~m_W$.}
 \label{money}
 \end{figure}
The RG equations for the  running of the Wilson coefficients, $c_T$ and $c_{WB}$, are  given by. 
\bea
{c}_T(m_W) = {c}_T(\Lambda)-\frac{1}{16\pi^2}  \log\left(\frac{\Lambda}{m_W}\right)\sum_i\gamma^i_{T}c_i \nonumber\\
{c}_{WB}(m_W) =  {c}_{WB}(\Lambda)-\frac{1}{16\pi^2} \log\left(\frac{\Lambda}{m_W}\right)\sum_i \gamma^i_{WB}c_i  
\label{rg}
\eea
where the anomalous dimensions, $\gamma^i_{T}$, were derived in Ref.~\cite{GuptaU} and are shown in Table~\ref{anadim}. As discussed in Sec.~\ref{intro}, it is the weakly constrained operators whose RG contributions can impact our analysis.  Because of the strong constraints on $c_{2B},c_{2W}, c_{BB}$ and $c_{WW}$  described in Sec.~\ref{bounds}, their RG contributions to $\hat{S}$ and  $\hat{T}$  are ${\cal O}(10^{-5})$ or smaller and thus have negligible impact on our analysis. Finally, we find that given the bound from Ref.~\cite{GrojeanM}, $|\lambda_\gamma|\lesssim 0.01$ ,  the RG contribution of $c_{3W}$ to  $c_{WB}$ also turns out to be small (of   ${\cal O}(10^{-4})$) and will  be neglected. 

 All the papers addressing the CDF anomaly with heavy new physics and using Fig.~\ref{green}, have the underlying assumption that all Wilson coefficients, including the weakly constrained ones,  are  at least as small as those generating $\hat{S}$ and  $\hat{T}$. Thus all the RG contributions in the right hand side of \eq{rg} can be ignored to yield $c_T(m_W) \approx {c}_T(\Lambda)$ and $c_{WB}(m_W) \approx  {c}_{WB}(\Lambda)$. The what has been considered so far are a subset of the full set of possibilities.

We now consider the RG effect of the less well constrained operators.
First, consider  the individual RG contribution due to $c_{H-}$ alone, 
\bea
\Delta \hat{S}=\frac{c_H v^2}{6 \Lambda^2} \frac{g^2}{16\pi^2}  \log\left(\frac{\Lambda}{m_W}\right),~~~~~~ \Delta \hat{T}=-\frac{3 c_H v^2}{2\Lambda^2} \frac{g'^2}{16\pi^2}  \log\left(\frac{\Lambda}{m_W}\right). \label{dS}
\eea
These contributions alone can give the necessary contribution to $\hat{S}$ and $\hat{T}$ to explain the anomaly.   For instance, taking $\Lambda=10~m_W$ we find that the anomaly can be explained  from the contributions  in \eq{dS} alone for   $-2.5\leq c_H\leq -1.4$. This range is also allowed by the bound on $\kappa_V$ in \eq{kappalimit}. Within the SMEFT framework, however, it is difficult to accommodate  $c_H<0$ on general grounds~\cite{rattazzi}.\footnote{Ref.~\cite{rattazzi} mentions that the only  exception, where this does not hold, is the case of models with scalar triplets.} Going beyond the SMEFT to scenarios where electroweak symmetry is non-linearly realised,   one can interpret the RG contributions in \eq{dS} as contributions due to a non-standard gauge-Higgs coupling, $\kappa_V \neq 1$. As discussed in Ref.~\cite{Cacci}, in such models it is possible to have  $\kappa_V>1$  thus making it possible to explain the anomaly; Ref.~\cite{Cacci}  also includes   a discussion of possible UV completions. As  this work assumes the SMEFT framework, however, we  take $0 \leq  c_{H-}\leq 0.64$, with the  upper limit arising from \eq{kappalimit}. In this range,  we find only  a small  $\Delta \hat{S}, \Delta \hat{T} \lesssim 10^{-4}$  that will  be neglected in our analysis.\footnote{Note that  strictly speaking $\kappa_V$  constrains  $c_{H-}(m_h)$ and not  $c_{H-}(\Lambda)$. However,   $c_{H-}(m_h)$ being amongst the  least constrained operators, all RG effects contributing to the difference, $(c_{H-}(\Lambda) -c_{H-}(m_h))$, are genuinely one-loop suppressed compared to $c_{H-}(\Lambda)$. Assuming $c_{H-}(\Lambda)=c_{H-}(m_h)$ thus only results in a two loop order correction to $\Delta \hat{S}$ and $\Delta \hat{T}$ in \eq{dS}.  }

 This still leaves us with the  four UV parameters: $c_{HB}(\Lambda), c_{HW}(\Lambda), c_{WB}(\Lambda)$ and $c_{T}(\Lambda)$. It will be convenient to use.,
 \bea
 \hat{S}_\Lambda=\frac{g^2 v^2 c_{WB}(\Lambda)}{\Lambda^2}~~~~~~~~~~~\hat{T}_\Lambda=\frac{v^2 c_{T}(\Lambda)}{\Lambda^2},
 \eea
i.e. the   UV contribution to   $\hat{S}$ and $\hat{T}$ from the matching scale, as two of our UV parameters instead of  $c_{WB}(\Lambda)$ and $c_{T}(\Lambda)$.  Using \eq{rg} it is possible to  derive an allowed region, in terms of these 4 UV parameters,  that corresponds to the green  area shown  in  Fig.~\ref{green}. That is to say,  the 2 dimensional constraints in Fig.~\ref{green}  can be RG evolved to a 4 dimensional hyperspace at the matching scale, $\Lambda$. As it is difficult to visualise this 4 dimensional space we show slices of $c_{HB}(\Lambda)$-$c_{HW}(\Lambda)$ planes for some fixed  values of $ \hat{S}_\Lambda$ and $ \hat{T}_\Lambda$ in Fig.~\ref{slices}. We also show in yellow the region allowed by the TGC bounds in Fig.~\ref{yellow} from Ref.~\cite{GrojeanM, cmswz}.  A resolution of  the CDF anomaly is possible only if there is an overlap between the green and yellow regions.   The first row shows three $c_{HB}(\Lambda)$-$c_{HW}(\Lambda)$ slices with  $\hat{T}_\Lambda=0$ but different  $\hat{S}_\Lambda=0$ values. The fact that there is an allowed region in the first two plots establishes one of the main claims of this work:   a non-zero  $\hat{T}_\Lambda$ at the matching scale is not required  to explain the anomaly.  For  the three figures in the  second row we take a much smaller   $\hat{T}_\Lambda$  than the value thought to be required from Fig.~\ref{green}. We find that the anomaly can be resolved in each of the three cases.

In Fig.~\ref{money} we show the region consistent with the CDF anomaly after marginalising over $c_{HB}(\Lambda)$ and $c_{HW}(\Lambda)$,  while obeying  the TGC bounds from Fig.~ \ref{yellow}. We find a much larger allowed region than what is obtained by simply putting   $c_{HB}(\Lambda) = c_{HW}(\Lambda) =0$ (also shown in Fig.~\ref{money}). This region includes vanishing and even negative values of $\hat{T}$. The shape of the region is easy to understand as the effect of non-zero values of  $c_{HB}(\Lambda)$ and $c_{HW}(\Lambda)$  is to shift the origin of the allowed region with respect to its position for  $c_{HB}(\Lambda) = c_{HW}(\Lambda) =0$.

It is worth emphasising that the scale of new physics, $\Lambda \sim 800$ GeV considered here  is much  smaller than the ${\cal O}$ (6 TeV) scale~\cite{Strumia} required if  UV contributions to the $\hat{S}$ and  $\hat{T}$ parameters are to entirely explain the anomaly. This is because while the latter contribute to the mass of the $W$ boson at tree level, the $c_{HB}, c_{HW}$ contributions considered here are loop suppressed.  This also implies large indirect effects in the form of TGC and anomalous Higgs couplings. In particular the operators , ${\cal O}_{HB}$ and ${\cal O}_{HW}$, are expected to be probed very precisely in the future due to their high energy contribution to processes like diboson production~\cite{wulzer, GrojeanM}, Higgsstrahlung~\cite{Banerjee1} and Higgs production via vector boson fusion (VBF)~\cite{Araz}. For instance notice  the dashed curve in Fig.~\ref{slices} where we show the HL-LHC projections for TGC measurements from Ref.~\cite{GrojeanM};  it is clear that much of the parameter space consistent with the CDF anomaly would be probed by the HL-LHC.  The relatively low scale also increases the prospect of a direct discovery if the new physics responsible for the anomaly.  

\subsection{Extending the analysis beyond the universal case} 
\label{nuniv}

We have restricted  our analysis to the case of universal new physics only for practical reasons as an analysis involving all the dimension 6 operators will be far too elaborate. We would like to emphasise however that the new regions in the allowed parameter space discovered in this  work, would remain even in the presence of additional non-universal operators. If anything, a completely general analysis may reveal  other  regions in the SMEFT  parameter space  also consistent with the new CDF $W$-mass measurement. Let us discuss why this is so. 

First, to perform a general analysis we will have to augment the 16 bosonic operators of Table~\ref{table1} by adding 43 dimension 6 operators (for a single generation of fermions, see Ref.~\cite{warsaw}) to obtain a complete 59 dimensional basis (see for eg. Ref.~\cite{GuptaU}). There would be two more operators  that give tree level contribution to the $W$-mass in addition to ${\cal O}_{WB}$ and  ${\cal O}_{T}$ (namely, the operators  ${\cal O}^{(3)}_{Hl}$ and  ${\cal O}_{ll}$ from Ref.~\cite{warsaw}, see Ref.~\cite{Bagnaschi:2022whn, Balkin:2022glu}). There would be many more operators contributing to these 4 operators via RG effects. A complete analysis should include all these operators and inspect which of them can have a large impact via their RG contributions. This can, in principle, result in the discovery of other allowed regions in the  SMEFT parameter space at  the matching scale. 

As far as the regions shown in Fig.~\ref{slices} is concerned, they will remain consistent with the CDF measurement. This is because these regions would still exist in the limit that the 43 new Wilson coefficients  vanish. If these new Wilson coefficients are marginalised over, the regions shown in in Fig.~\ref{slices} will only become larger. 
\begin{table}
\begin{center}
\begin{tabular}{C{0.8cm}|c c}
 & $c_{H-}$  & $c_{T}$ \\
\hline 
$\gamma^i_{c_{T}}$  & $\frac{3}{2}g^{\prime2}$  & $\frac{9}{2}g^{2}+12\lambda+12y_{t}^{2}$ \\
$\gamma^i_{c_{WB}}$  & $-\frac{1}{6}$ & $-\frac{1}{2}$ \\
\end{tabular}

\vspace{0.5cm}

\begin{tabular}{C{0.7cm} | C{3.115cm} C{3.115cm} C{3.115cm} C{3.115cm} }
 & $c_{2B}$  & $c_{2W}$  & $c_{HB}$  & $c_{HW}$ \tabularnewline
\hline 
$\gamma^i_{c_{T}}$  & $3g^{\prime4}+\frac{9}{8}g^{\prime2}g^{2}+3\lambda g^{\prime2}$  & $\frac{9}{8}g^{\prime2}g^{2}$  & $-\frac{9}{4}g^{\prime2}g^{2}-6\lambda g^{\prime2}$ & $-\frac{9}{4}g^{\prime2}g^{2}$\tabularnewline
$\gamma^i_{c_{{{WB}}}}$  & $\frac{1}{4}\left( \frac{147}{8}-\frac{53}{4}t_{\theta_{W}}^{2}\right)g^{\prime2}$ & $ \frac{77}{32}g^{2}+\frac{29}{16}g^{\prime2}$ & $-\frac{9}{8}g^{2}-\frac{7}{24}g^{\prime2}-\lambda$ & $\frac{5}{8}g^{2}+\frac{1}{8}g^{\prime2}-\lambda$\tabularnewline
\end{tabular}

\vspace{0.5cm}

\begin{tabular}{C{0.8cm} | C{3.115cm} C{3.115cm} C{3.115cm} C{3.115cm} }
 & $c_{BB}$  & $c_{WW}$ & $c_{WB}$ & $c_{3W}$ \tabularnewline
\hline 
$\gamma^i_{c_{T}}$  & $0$  & $0$  & $0$  & $0$ \tabularnewline
$\gamma^i_{c_{WB}}$  & $2g^{\prime2}$  & $2g^{2}$  & $\frac{9g^{2}}{2}-\frac{g^{\prime2}}{2}+6y_{t}^{2}+4\lambda$  & $-\frac{g^{2}}{2}$ \tabularnewline
\end{tabular}
\caption{\small Anomalous dimension matrix for the Wilson coefficients $c_T$ and $c_{WB}$ of the dimension-6 bosonic operators, in the HISZ~\cite{hisz}. }
\label{anadim}
\end{center}
\end{table}

\section{Conclusions}

In this work we have discussed some important aspects about the SMEFT  interpretation of the CDF $W$-mass anomaly. In particular we have carried  out an RG evolution of the SMEFT parameter space  consistent with the anomaly from the scale, $m_W$, to  the matching scale. We find  that the matching scale  SMEFT parameter space capable of explaining the anomaly is much larger owing to these RG effects. We show this parameter space in Fig.~\ref{money} where  it is interesting to note that there can be solution to the anomaly with a vanishing or even negative $T$-parameter. The larger parameter space at the SMEFT level will hopefully lead to the identification of new UV complete models that have not been considered yet. 

This large impact of RG effects arises due to the running  of the operators, ${\cal O}_{HB}$ and ${\cal O}_{HW}$ in Table~\ref{table1}-- that are only weakly constrained by diboson measurements --to the much more strongly constrained operators  that generate the   $S$ and $T$-parameters. Fig.~\ref{slices} shows the parameter space required to explain the CDF anomaly in the plane of these two operators. 

As the new regions of SMEFT parameter space, discovered here, arise due to one-loop RG effects to the $W$-mass from the above operators, for them to be significant requires a relatively scale ($\Lambda~\sim 800$ GeV). The scale  required in the case where the $W$-mass anomaly arises only from UV contributions to the $S$ and $T$ parameters is much higher, around a few TeV.  This opens the possibility of  discovering the new physics responsible for the anomaly in future direct searches as well as via indirect probes in the diboson, Higgstrahlung and VBF  channels.

While this analysis has been  performed in the context  of the CDF $W$-mass anomaly, we  hope that the approach developed in this work can find more general applicability for SMEFT studies.

\bibliographystyle{JHEP}
\bibliography{references}    
\end{document}